\documentclass[twoside]{ilcws08}
\usepackage[latin1]{inputenc}
\usepackage[dvips]{graphicx,epsfig,color}
\usepackage{wrapfig,rotating}
\usepackage{amssymb,amsmath,array}

\begin{document}
\title{
%%%%   Paper title goes here  %%%%%%%%%%%%%%
LiC Detector Toy - Tracking detector optimization with 
fast simulation and its application to the ILD design} %% 
%***********************************************************************
% AUTHORS INFORMATION AREA
%***********************************************************************
\author{{\sl Manfred Valentan, Meinhard Regler, Winfried Mitaroff} and {\sl Rudolf Fr\"uhwirth}
% Optional short acknowledgment: remove next line if non-needed
%\thanks{This is an optional funding source acknowledgment.}
% DO NOT MODIFY THE FOLLOWING '\vspace' ARGUMENT
\vspace{.3cm}\\
% Addresses and institutions (remove "1- " in case of a single institution)
Austrian Academy of Sciences -- Institute of High Energy Physics \\
Nikolsdorfer Gasse 18, A-1050 Vienna, Austria, EU
%% Remove the next three lines in case of a single institution
%\vspace{.1cm}\\
%2- Second Author's Institution - Department \\
%Address of Second Author's Institution - Country\\
}
%%***********************************************************************
% END OF AUTHORS INFORMATION AREA
%***********************************************************************

\maketitle

\begin{abstract}
The ``LiC Detector Toy'' is a fast single-track simulation and reconstruction tool, aiming at the optimization of tracking detector design, i.e. geometric layout and material budgets. Its implementation is based on the {\sc MatLab} system.
Improvements over the last year include correct handling of complex forward regions (arbitrary mixture of cylindrical and plane surfaces), enhanced detector description, flexible data presentation of results (e.g. fitted track resolutions and impact parameters), and support by an integrated GUI. In addition a non-GUI version running under open-source {\sc Octave} has been implemented.
The tool has recently been used for fixing the ``reference design'' layout of the silicon tracker (SIT, SET and FTD) of the ILD detector concept.
\end{abstract}

% \section{The ``LiC Detector Toy'' program}
\section{Simulation and reconstruction}

The ``LiC Detector Toy'' (LDT) -- a simple but powerful and flexible software tool for detector design, modification and upgrade studies -- aims at investigating the resolution of reconstructed track parameters for comparing and optimizing the track-sensitive devices and material budgets of various detector layouts. 
This is achieved by a simplified simulation of the investigated setup, followed by full single track reconstruction. 

% \subsection{Detector model}
The detector model corresponds to a collider experiment with a solenoid magnet, 
and is rotationally symmetric w.r.t. the beam axis (denoted $z$); 
the surfaces are either cylinders (``barrel region'') or planar disks (``forward/rear region''). 
The magnetic field $\vec{B} = (0, 0, B_z)$ is homogeneous and parallel to the $z$-axis, 
thus implying a helix track model (radius $R_{\mathrm{H}}$). 
All material causing multiple scattering is assumed to be concentrated within thin layers; 
it can either be averaged over a surface, or can be modeled individually. 

%\begin{wrapfigure}{r}{0.4\columnwidth}
%\centerline{\includegraphics[width=0.3\textwidth]{detector_display_3D.eps}}
%\includegraphics{mainwindow.jpg}
%\caption{ILD detector model}\label{fig:detectormodel}
%\end{wrapfigure}

% \subsection{Simulation}
Simulation generates charged tracks with user-defined ranges for the start parameters (vertex position, momentum $p$ or its transverse component $p_{\mathrm{T}}$, and polar angle $\vartheta$);
performs helix tracking with breakpoints due to multiple scattering according to 
Rossi-Greisen-Highland;\footnote{
Energy loss due to bremsstrahlung of $e^{\pm}$ will be included in a future release.}
and simulates detector measurements including systematic or stochastic inefficiencies, and uniform or Gaussian observation errors, {\it but no other degradation of data}.
At present supported is any combination of $Si$ strip detectors (single or double sided, with any stereo angle $\alpha$), pixel detectors, and a time projection chamber (TPC).

% \subsection{Reconstruction}
Reconstruction is done separately for each track from all its simulated measurements 
{\it without pattern recognition}, by fitting the five parameters and $5 \times 5$ covariance matrix at a pre-defined reference cylinder, e.g. the inner surface of the beam tube. 
Standard parameters are [$\Phi, z, \vartheta, \beta = \varphi - \Phi, \kappa = \pm 1/R_{\mathrm{H}}]$ 
($\Phi$ and $\varphi$ denote the azimuth angles of position and direction, respectively).
% They  may optionally be converted to a Cartesian space point \& momentum representation. 
The method used is an inwards-running Kalman filter~\cite{cup00,kalman}, with linear expansion from a ``reference track'' being defined by the simulation; 
process noise is calculated from the material budget as seen by that track. 
Sensitive tests of goodness of the fits ($\chi^2$ distributions and pull quantities) 
are an integral part. 

% \subsection{Algorithms used}
The algorithms used in the tool stand on solid mathematical ground and are field-proven. 
The software is based on {\sc MatLab}\textsuperscript{\textregistered} \cite{matlab}, a licenced high-level matrix algebra language and IDE; usage is supported by a graphic user interface (GUI). Zoomable 2D and 3D displays of the setup and flexible data presentation of the results are easily at hand. 
The core program has recently been refactored to run also in command line mode without GUI under {\sc Octave} \cite{octave}, an open-source product which is source-code compatible.

The tool is deliberately kept simple and can without effort be adapted to individual needs. Its main purpose, however, is to supply a tool for non-experts without algorithmic knowledge. 
Once the detector description (``input sheet'') has been set up, individual detector layers can easily be added, moved, modified or removed, and the result of these changes can quickly be evaluated. For more information see the User's Guide \cite{lictoy}.

\section{Application to the ILD design}

A first ``ILD reference design'' was proposed in Sept.~2008 at Cambridge (UK); however, details of the Silicon Tracker layout were left open. Optimization brainstormings, based on LDT v2.0 and validated against full simulation (Mokka--Marlin), took place in Sept/Oct.~2008 at DESY, and in Dec.~2008 at the SiLC meeting in Santander. They helped defining the layout eventually used by the benchmark MC runs for the LoI.

\subsection{Barrel region (SIT and SET):}
The Silicon Inner Tracker (SIT) is placed between Vertex Detector and TPC, and the Silicon External Tracker (SET) fills the gap between TPC and Electromagnetic Calorimeter. 
In order to assess the impact of SIT and SET on the overall tracking performance, the standard layout is compared with two alternatives by removing either SIT or SET, but not both,
for radial tracks ($\vartheta=90^{\circ}$) with momenta $p = p_{\mathrm{T}} = 1 \ldots 1000$~GeV/$c$.

\begin{figure}[ht]
  \begin{center}
    \includegraphics[width=0.9\textwidth]{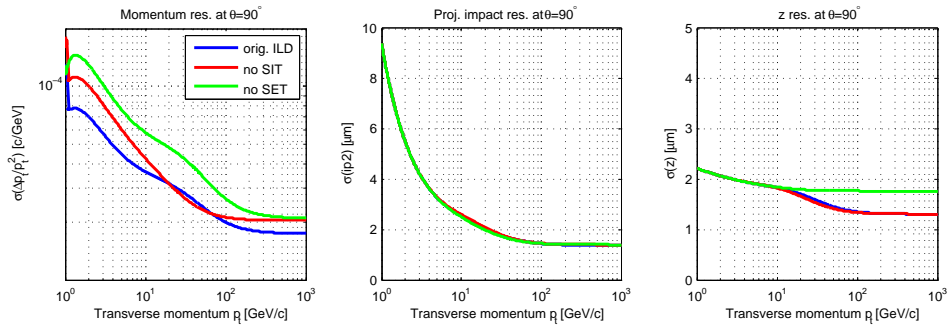}
    \caption{{\sl Effects of removing either SIT (red) or SET (green) w.r.t. standard layout (blue).}}
    \label{fig:effectSITSET}
  \end{center}
\end{figure}

Figure~\ref{fig:effectSITSET} shows, for these 3 layouts, 
the resolutions of $\Delta p_{\mathrm{T}} / p_{\mathrm{T}}^2$ (left), transverse impact parameter (middle), and $z$ coordinate (right) as functions of $p_{\mathrm{T}}$.

\vspace{-0.5cm}
Note from the right figure that the SET also contributes to the $z$ coordinate resolution because of its precise $z$ measurement at long lever arm.

\subsection{Forward/rear region (FTD):}
The Forward Tracking Detector (FTD) is placed between beam tube and inner wall of the TPC. 
It consists, in each $z$ direction, of 3 pixel disks (FTD~1--3) and 4 strip disks (FTD~4--7), the latter double sided with stereo angles $\alpha = 90^{\circ}$.
This standard setup is compared with three alternatives which make use of strip disks with a small stereo angle $\alpha = 6^{\circ}$, while leaving everything else unchanged:

\begin{center}
\begin{tabular}[b]{| l | c | r r r r | l |} \hline
Layout      & symbolic & FTD 4           & FTD 5           & FTD 6                 & FTD 7           \\
\hline \hline
Standard  & + + + +    & $90^{\circ}$ & $90^{\circ}$ &      $90^{\circ}$ & $90^{\circ}$ \\
Setup 1    & II II + II     & $   6^{\circ}$ & $  6^{\circ}$ &       $90^{\circ}$ & $  6^{\circ}$ \\
Setup 2    & II II II +     & $   6^{\circ}$ & $  6^{\circ}$ &       $  6^{\circ}$ & $90^{\circ}$ \\
Setup 3    & II II = II     & $   6^{\circ}$ & $  6^{\circ}$ & \P~$  6^{\circ}$ & $  6^{\circ}$ \\
\hline
\end{tabular}
\footnote{
Normally the strips of one side are oriented radially, measuring the azimuthal arc $R \Phi$, and the strips of the other side are inclined by $\alpha$. \P~Only FTD~6 of setup~3 has its strips of one side oriented tangentially, measuring the radius $R$.
In case of $\alpha = 90^{\circ}$, this distinction becomes dummy.}
\end{center}

\begin{figure}[ht]
  \begin{center}
    \includegraphics[width=0.9\textwidth]{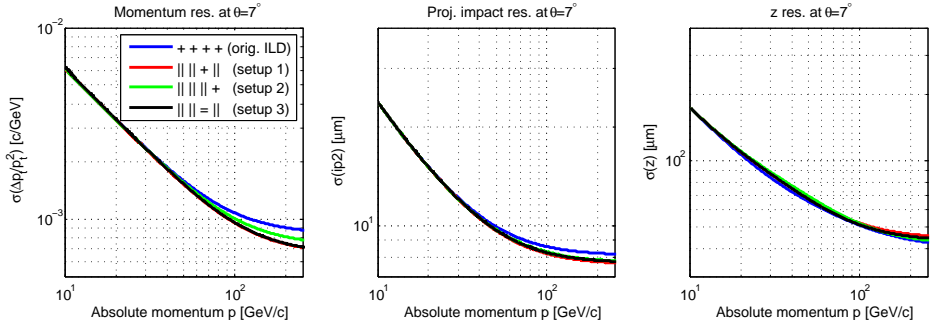}
    \caption{{\sl 
       Alternative setups 1 (red), 2 (green) and 3 (black) vs. standard setup (blue).}}
    \label{fig:optimizationforward}
  \end{center}
\end{figure}

The impact on the tracking performance of these 4 setups are tested with forward tracks ($\vartheta=7^{\circ}$) of momenta $p = 10 \ldots 250$~GeV/$c$.
Figure~\ref{fig:optimizationforward} 
shows the resolutions of $\Delta p_{\mathrm{T}} / p_{\mathrm{T}}^2$ (left), transverse impact parameter (middle), and $z$ coordinate (right) as functions of $p$.

\vspace{-0.5cm}
Thus, it is possible to achieve better resolutions in transverse momenta and impact parameters at high momenta, at the cost of only small losses in the resolution of $z$.

%\section{Bibliography}
 
% ****************************************************************************
% BIBLIOGRAPHY AREA
% ****************************************************************************

\begin{footnotesize}
% IF YOU DO NOT USE BIBTEX, USE THE FOLLOWING SAMPLE SCHEME FOR THE REFERENCES
% ----------------------------------------------------------------------------

% IF YOU USE BIBTEX,
% - DELETE THE TEXT BETWEEN THE TWO ABOVE DASHED LINES
% - UNCOMMENT THE NEXT TWO LINES AND REPLACE 'Name_Of_Your_BibFile'

%\bibliographystyle{unsrt}
%\bibliography{Name_Of_Your_BibFile}
% example of Name_Of_Your_BibFile.bib
% @Article{Turcato:2006ch,
%      author    = "Turcato, M.",
%  collaboration = "ZEUS and H1",
%      title     = "Lepton flavour violation and charmonium physics at HERA",
%      journal   = "Nucl. Phys. Proc. Suppl.",
%      volume    = "162",
%      year      = "2006", 
%      pages     = "283-287",
%      SLACcitation  = "%%CITATION = NUPHZ,162,283;%%"
% }
% 
% @Unpublished{Gogitidze:2007du,
%      author    = "Gogitidze, N.",
%  collaboration = "H1", 
%      title     = "Prompt photons and particle momentum distributions at
%                   HERA", 
%      year      = "2007",
%      note    = "hep-ex/0701033",
%      SLACcitation  = "%%CITATION = HEP-EX 0701033;%%"
% }

\end{footnotesize}

% ****************************************************************************
% END OF BIBLIOGRAPHY AREA
% ****************************************************************************

\end{document}